\documentclass[aps,prb,twocolumn,groupedaddress,amsfonts,showpacs]{revtex4}
\usepackage{graphicx,color}
\usepackage{bm}
\usepackage[latin1]{inputenc}

\begin{document}

\title{Electron-electron interaction in 2D and 1D ferromagnetic (Ga,Mn)As}

\author{D. Neumaier}
\email{daniel.neumaier@physik.uni-regensburg.de}
\author{M. Schlapps}
\author{U. Wurstbauer}
\author{J. Sadowski}
\altaffiliation[Present address: ]{MAX-Laboratory, Lund University,
SE-221 00, Lund, Sweden.}
\author{M. Reinwald}
\author{W. Wegscheider}
\author{D. Weiss}

\affiliation{Institut f\"{u}r Experimentelle und Angewandte Physik,
Universit\"{a}t Regensburg, 93040 Regensburg, Germany}

\date{\today}

\begin{abstract}
We investigated the magnetotransport in high quality ferromagnetic
(Ga,Mn)As films and wires. At low temperature the conductivity
decreases with decreasing temperature without saturation down to 20
mK. Here we show, that the conductivity decrease follows a
ln($T/T_0$) dependency in  2D films and a $-1/\sqrt{T}$ dependency
in 1D wires and is independent of an applied magnetic field. This
behavior can be explained by the theory of electron-electron
interaction.

\end{abstract}


\maketitle

Up to now the ferromagnetic semiconductor (Ga,Mn)As \cite{Ohno} is
one of the best understood ferromagnetic semiconductors and has
become a promising candidate for future spintronic devices. The
Mn ions substituting Ga on the regular sites of the zinc-blende
lattice provide both, holes and magnetic moments. The ferromagnetic
order between the individual Mn ions is mediated by these holes
\cite{Dietl}. Ferromagnetism in (Ga,Mn)As is well understood,
allowing to predict Curie temperature \cite{Dietl},
magnetocrystalline anisotropies \cite{Sawicki} as well as
anisotropic magnetoresistance effects \cite{Baxter}. However, the
temperature dependence of the conductivity of (Ga,Mn)As is still
under debate. Starting at room temperature the conductivity decreases
with decreasing temperature until a local minimum is reached around
the Curie-temperature. Attempts to explain this minimum include e.g.
the formation of magnetic polarons \cite{Majumdar,Sawicki2}, or an
interplay with universal conductance fluctuations \cite{Timm}. A
recent work \cite{Moca} explains the temperature dependent
conductivity above and below $T_C$ within a picture invoking
localization effects using an extended version of the scaling theory
of Abrahams \textit{et al.} \cite{Abrahams,Zarand}. At temperatures
below $T_C$ the conductivity increases again in metallic samples,
reaches a local maximum at about 10 K before it drops again for
decreasing temperatures. The temperature dependence in this low
temperature regime is in the focus of the present communication. Attempts
to explain this behavior are, e.g. based on Kondo scattering
\cite{He}, weak localization \cite{Matsukura} or Mott-hopping
\cite{Esch}. A very recent report investigating 3D (Ga,Mn)As films
ascribes the conductivity decrease to Aronov-Altshuler scaling in 3D
\cite{Honolka}. By using one-dimensional (1D) and two-dimensional
(2D) samples we show below that the decreasing conductivity in this
regime can be ascribed to electron-electron interaction (EEI). The
effect of electron-electron interaction arises from a modified
screening of the Coulomb potential due to the diffusive propagation
\cite{Lee}. The expected temperature dependence of the conductivity
change for EEI depends on the dimensionality and goes with
ln($T/T_0$) for 2D-samples and with $-1/\sqrt{T}$ for 1D samples.
This behavior has been found in our experiments.

For the experiment we used three different (Ga,Mn)As wafers,
labeled, 1 - 3 (see Tab. 1), grown by low temperature molecular beam
epitaxy on semi-insulating GaAs \cite{Reinwald}. The nominal Mn
concentration of the (Ga,Mn)As layers varied between 4 \%  and 5.5
\% with corresponding Curie temperatures between 90 K and 150 K. To
investigate the transport properties of two- and one-dimensional
(Ga,Mn)As devices we fabricated  Hall-bars and arrays of wires using
optical lithography, e-beam lithography and subsequent reactive ion
etching. Arrays of wires were used to suppress universal conductance
fluctuations by ensemble averaging. Au contacts to the devices were
made by lift-off technique. The relevant parameters of the
investigated samples are listed in Tab. 1. Magnetotransport
measurements were carried out in a top loading dilution refrigerator
with a base temperature of 15 mK using standard 4-probe lock-in
techniques. Small measuring currents (25 pA to 1 nA) were used to
avoid heating.

\begin{table}
\begin{tabular}[b]{|l|p{0.8cm}|p{0.8cm}|p{0.8cm}|p{0.8cm}|p{0.8cm}|p{0.8cm}|p{0.8cm}|} \hline
Sample & $1_{2\mathrm{D}}$-a & $1_{2\mathrm{D}}$-b & $2_{2\mathrm{D}}$ & $3_{2\mathrm{D}}$ & $1_{1\mathrm{D}}$-a & $1_{1\mathrm{D}}$-b & $1_{1\mathrm{D}}$-c\\
\hline
\emph{L} ($\mu$m) & 180 & 180 & 60 & 180 & 7.5 & 7.5 & 7.5\\
\emph{w} ($\mu$m) & 11 & 11 & 7.2 & 10 & 0.042 & 0.042 & 0.035\\
\emph{t} (nm) & 42 & 42 & 20 & 50 & 42 & 42 & 42\\
\emph{N} & 1 & 1 & 1 & 1 & 25 & 25 & 12\\
\emph{a} (hours) & 0 & 51 & 8.5 & 0 & 0 & 51 & 0\\
$T_C$ (K) & 90 & 150 & 95 & 90 & 90 & 150 & 90\\
\emph{n} ($10^{26}/$m$^3$) & 3.8 & 9.3 & 1.7 & 3.1 & 3.8 & 9.3 & 3.8\\
\emph{$\rho$} ($10^{-5}$ $\Omega$m) & 3.5 & 1.8 & 13 & 5.2 & 3.5 & 1.8 & 3.5\\
\emph{$F^{2D}$, $F^{1D}$}  & 2.4 & 2.4 & 1.8 & 2.6 & 0.76 & 0.72 & 0.80\\
\hline
\end{tabular}
\caption{Length \emph{L}, width \emph{w}, thickness \emph{t} and
lines parallel \emph{N} of the samples. Some of the samples were
annealed for the time \emph{a} at 200 °C. Curie temperature $T_C$,
resistivity $\rho$ and carrier concentration \emph{n} were taken on
reference samples from the corresponding wavers. The screening
factor $F^{2D}$ and $F^{1D}$ were calculated using Eq. 1 and Eq. 2,
respectively. Different samples from the same waver are labeled a, b
or c.}
\end{table}

To investigate magnetotransport at milikelvin temperatures, we
measured the resistance of quasi 2D films and quasi 1D wires in a
perpendicular applied magnetic field. According to Lee \textit{et
al.} \cite{Lee} a (Ga,Mn)As film is considered to be two-dimensional
in the context of electron-electron interactions if the film
thickness $t$ is smaller than the thermal diffusion length
$l_T=\sqrt{\hbar D/k_BT}$. For our samples $l_T$ is $\sim200$ nm at
20 mK. Similarly, a sample is one-dimensional if both, wire width
$w$ and the wire thickness $t$, are smaller than $l_T$.

\begin{figure}
\includegraphics[width=\linewidth]{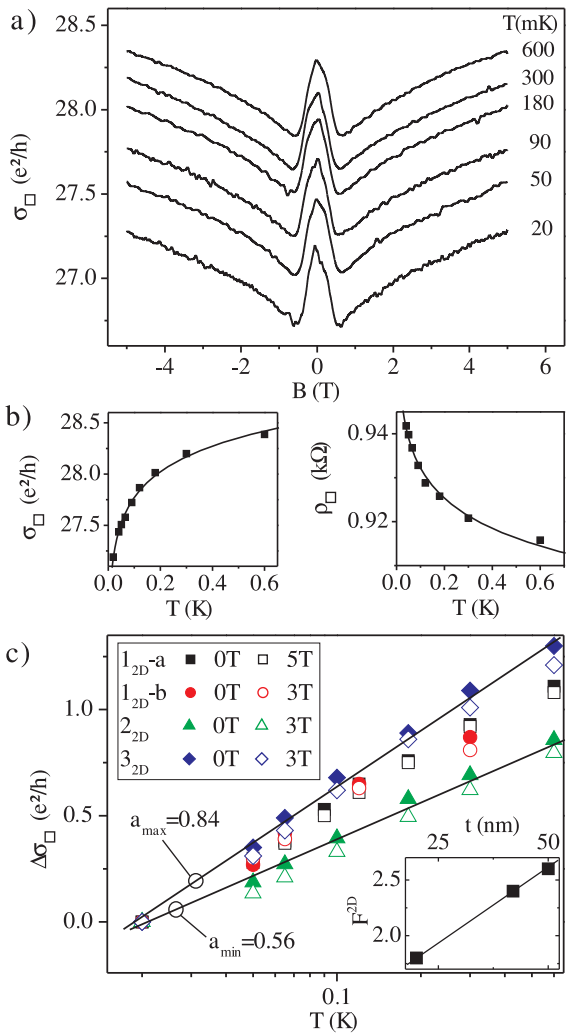}\\
\caption{a) Square conductivity of sample $1_{2\mathrm{D}}$-a at
different temperatures in a perpendicular applied magnetic field. b)
To the left: Square conductivity at $B=0$ for different
temperatures. To the right: Corresponding square resistivity at
$B=0$ for different temperatures. The straight lines are the best
ln($T/T_0$) fits. c) Change in square conductivity of the
investigated 2D samples relative to 20 mK, taken at zero magnetic
field (solid symbols) and $B=3$ T or $B=5$ T (open symbols). The
straight lines give a slope of $a_{max}=0.84$ and $a_{min}=0.56$,
which are the best linear fits for sample $2_{2\mathrm{D}}$ and
sample $3_{2\mathrm{D}}$, respectively. The inset shows the
screening factor $F^{2D}$ versus the (Ga,Mn)As layer thickness
\emph{t}.}
\end{figure}

We start with discussing temperature dependent transport for 2D
samples (Sample $1_{2\mathrm{D}}$-a, $1_{2\mathrm{D}}$-b,
$2_{2\mathrm{D}}$ and $3_{2\mathrm{D}}$ in Tab. 1). The discussion
for the 2D samples will be in terms of the square conductivity
$\sigma_{\Box}=\sigma\cdot t$, as the conductivity corrections due
to electron-electron interaction is expected to be independent on
the absolute value of $\sigma_{\Box}$. The conductivity $\sigma$ was
obtained by inverting the resistivity $\rho=R\cdot (t\cdot w)/l$,
with the resistance $R$ of the investigated samples: $\sigma=1/\rho$
The square conductivity of sample $1_{2\mathrm{D}}$-a is displayed
in Fig. 1a for temperatures between 20 mK and 600 mK in a
perpendicular applied magnetic field $B$. The shape of
$\sigma_{\Box}(B)$ remains unchanged in this temperature range and
is typical for ferromagnetic (Ga,Mn)As. The square conductivity
maxima at B = 0 stems from the anisotropic magnetoresistance (AMR)
\cite{Baxter} and reflects the fact that $\sigma_{\Box}$ for an
in-plane magnetization is higher than for an out-of-plane
magnetization. The positive slope of $\sigma_{\Box}$ for $B>400$ mT,
known as negative magnetoresistance (NMR), is discussed in terms of
increased magnetic order \cite{Nagaev}, or as a consequence of weak
localization in 3D \cite{Matsukura}. With decreasing temperature the
square conductivity decreases without saturation. This behavior is
also reflected in the temperature dependence of $\sigma_{\Box}$ and
the corresponding square resistance $\rho_{\Box}=\rho/t$ for zero
magnetic field which are displayed in Fig. 1b, respectively. This
behavior is in the focus of this communication. The square
conductivity change of sample $1_{2\mathrm{D}}$-a at different
temperatures relative to $\sigma_{\Box}$ measured at 20 mK is shown
in Fig. 1c for zero magnetic field (filled squares) and $B=5$ T
(open squares). For the temperature dependence of $\sigma_{\Box}$
due to EEI one obtains for a two-dimensional system \cite{Lee}:

\begin{equation}
\Delta
\sigma_{\Box}=\frac{F^{2D}}{\pi}\frac{e^2}{h}\log{\frac{T}{T_0}},
\end{equation}

with a screening factor $F^{2D}$, the electron charge $e$ and the
Planck constant $h$. The square conductivity change observed
experimentally follows such a logarithmical temperature dependency,
independent of the applied magnetic field, as it is expected from
the theory of electron-electron interaction. The slope in this
log($T$)-plot is 0.77. This corresponds to a screening factor
$F^{2D}$ of 2.4 (Eq. 1), which is also close to the screening factor
found in Co-films ($F^{2D}$(Co) = 2.0...2.6 \cite{Brands,Brands2}).

Does annealing of (Ga,Mn)As change the conductivity correction? Low
temperature annealing of (Ga,Mn)As causes an out-diffusion of
Mn-Ions from interstitial sites of the lattice, where they act as
double donors \cite{Edmonds}. Hence low temperature annealing
increases carrier concentration, square conductivity and
Curie-temperature \cite{Edmonds}. The relevant parameters of the
annealed sample $1_{2\mathrm{D}}$-a are listed in Tab. 1 (sample
$1_{2\mathrm{D}}$-b). Both carrier concentration and square
conductivity increased by a factor of approx. 2 after annealing. The
screening factor $F^{2D}$, describing the strength of EEI, however,
remained unchanged, as is shown in Fig. 1c. This demonstrates that
the observed conductivity decrease is a universal phenomenon, as it
is independent on the absolute value of $\sigma_{\Box}$. A recent
experiment of He \textit{et al.} \cite{He} seems to be in contrast
to our finding. These authors observed a reduction of the
logarithmical slope of the resistivity due to low temperature
annealing. If we plot for our samples instead of $\sigma_{\Box}$ the
square resistivity change versus log($T$) we obtain the same result.
The change in square resistivity $\Delta \rho_{\Box}$ is connected
to the change in square conductivity $\sigma_{\Box}$ by: $\Delta
\rho_{\Box}=1/\sigma_{\Box1}-1/\sigma_{\Box2}\approx
\rho_{\Box}^2\Delta \sigma_{\Box}$. Though $\Delta \sigma_{\Box}$
remains unchanged by annealing, $\Delta \rho_{\Box}$ does not. As
$\rho_{\Box}$ is decreasing due to annealing, also $\Delta
\rho_{\Box}$ is decreasing. From the data of He \textit{et al.}
\cite{He} we estimate an average screening factor $F^{2D}$ of 2.5.
This is in good agreement with our results. A monotonic dependency
of the logarithmical slope and so of the screening factor $F^{2D}$
with annealing can not be found.

The square conductivity change of all investigated 2D samples follows a
logarithmical temperature dependency and is independent of an
applied magnetic field (Fig. 1c). The screening factor $F^{2D}$ of
the investigated 2D samples depends on the layer thickness $t$.
$F^{2D}$ vs. $t$ is plotted in the lower inset of Fig. 1c. For the
investigated 4 samples the screening factor scales linearly with the
layer thickness $t$. This behavior suggests, that screening might
play a role, although the Thomas-Fermi screening length is smaller
than 1 nm and by this much smaller than the layer thickness $t$ of
the investigated samples (20 nm- 50 nm).

To investigate the conductivity correction in one dimension we
fabricated nanowire arrays to suppress aperiodic conductance
fluctuations by ensemble averaging. A corresponding electron
micrograph of sample $1_{1\mathrm{D}}$-a with 25 nanowires connected
in parallel is shown in Fig. 2a. Each wire has a length $l$ of 7.5
$\mu$m and a width $w$ of 42 nm. The magnetoconductance of sample
$1_{1\mathrm{D}}$-a is shown in Fig. 2b for different temperatures
between 20 mK and 600 mK. Also for 1D samples the conductance
decreases with decreasing temperature without saturation. While for
$T > 50$ mK the shape of the magnetoconductance is affected by AMR
and NMR , similar to the 2D samples discussed above, a weak
localization correction has been found at the lowest temperatures.
This weak localization in 1D was discussed in detail in Ref.
\onlinecite{WAL}. In this communication we focus on temperatures
above 50 mK, were no weak localization can be observed.

\begin{figure}
\includegraphics[width=0.85\linewidth]{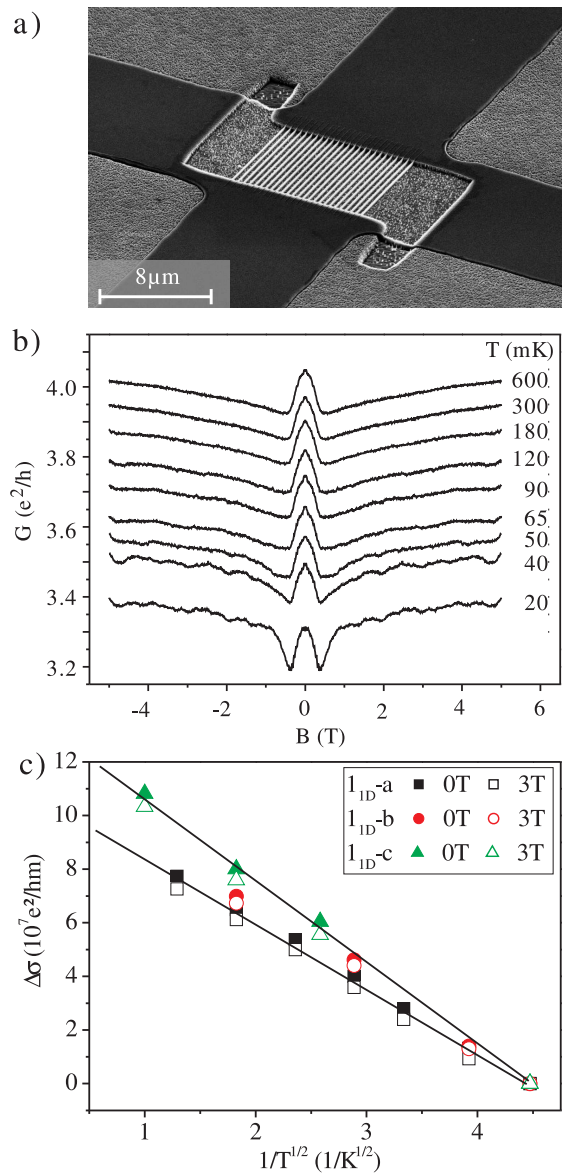}\\
\caption{a) Electron micrograph of sample $1_{1\mathrm{D}}$-a having
25 wires in parallel. b) Magnetoconductance of sample
$1_{1\mathrm{D}}$-a in a perpendicular applied magnetic field at
different temperatures. c) Conductivity change of the investigated
1D samples relative to 50 mK, taken at \emph{B}=0 (solid symbols)
and \emph{B}=3 T (open symbols). The solid lines are the best linear
fits for sample $1_{1\mathrm{D}}$-a and $1_{1\mathrm{D}}$-c.}
\end{figure}

The conductivity change of three 1D samples, taken relative to the
conductivity at 50 mK and at zero magnetic field (filled symbols)
and 3 T (open symbols), is shown in Fig. 2c. The data points follow
a $1/\sqrt{T}$ dependency independent of the applied magnetic field.
Such a dependence is expected for EEI correction in 1D which is
given by \cite{Lee}:

\begin{equation}
\Delta \sigma=-\frac{F^{1D}}{\pi
A}\frac{e^2}{\hbar}\sqrt{\frac{\hbar D}{k_BT}},
\end{equation}

with a screening factor $F_{1D}$, the wire cross section $A$, the
diffusion constant $D$ and the Boltzmann constant $k_B$. Again, the
conductivity decrease with decreasing $T$ can also be explained in
terms of electron-electron interaction. The relevant screening
factor $F_{1D}$ (Eq. 2) ranges between 0.72 and 0.80. Also the value
of $F_{1D}$ is close to the value found in Ni-wires ($F_{1D}$(Ni) =
0.83 \cite{Ono}).

Our results obtained from 1D samples and 2D samples show, that
electron-electron interaction is dominating the conductance of
(Ga,Mn)As in the low temperature regime. Comparing our result with
recent results of Honolka \textit{et al.} \cite{Honolka}, the most
obvious difference is the resistivity of the investigated samples.
While the resistivity of the samples investigated in Ref.
\onlinecite{Honolka} is $\sim7\cdot10^4$ $\Omega$m at 300 mK, the
resistivity of our samples is between 5.5 times (sample
$2_{2\mathrm{D}}$) and 40 times lower (sample $1_{2\mathrm{D}}$-b).
While the devices investigated by Honolka \textit{et al.} are
already close to a metal-insulator transition, where the conductance
can be described by Altshuler-Aronov scaling, we are still on the
metallic side of conductance, were the conductance is dominated by
electron-electron interaction. The good agreement of screening
factors in 1D and 2D (Ga,Mn)As samples with those found in
conventional metallic ferromagnets like Co \cite{Brands,Brands2} and
Ni \cite{Ono} underlines, that we are on the metallic side of
conductance. The mechanisms causing the conductivity decrease in
both ferromagnetic semiconductor (Ga,Mn)As and ferromagnetic metals
seem to be very similar. A comparable result was found by Maliepaard
\textit{et al.} in n-doped GaAs \cite{Maliepaard}. While on the
metallic side conductance was dominated by electron-electron
interaction, close and beyond the metal insulator transition the
conductance could be described by Altshuler-Aronov scaling.

In summary we have shown that the conductivity of (Ga,Mn)As
decreases with decreasing temperature below 1 K. The observed
conductivity decrease in wires and films on the metallic side of
conductance can be described by EEI. The observed screening factors
$F_{1D}$ and $F_{2D}$ are in good agreement with the screening
factors found in conventional metallic ferromagnets.

Acknowledgement: We thank the German Science Foundation (DFG) for financial support via SFB 689.




\begin{thebibliography}{00}

\bibitem{Ohno} H. Ohno, Science \textbf{281}, 951 (1998).

\bibitem{Dietl} T. Dietl, H. Ohno, F. Matsukura, J. Cibert, and D. Ferrand, Science
\textbf{287}, 1019 (2000).

\bibitem{Sawicki} M. Sawicki, J. Magn. Magn. Mater. \textbf{300}, 1 (2006) and references therein.

\bibitem{Baxter} D. V. Baxter, D. Ruzmetov, J. Scherschligt, Y. Sasaki, X. Liu, J. K.
Furdyna, and C. H. Mielke, Phys. Rev. B \textbf{65}, 212407 (2002).

\bibitem{Majumdar} P. Majumdar, and P. B. Littlewood, Nature (London) \textbf{395}, 479 (1998).

\bibitem{Sawicki2} M. Sawicki, T. Dietl, J. Kossut, J. Igalson, T. Wojtowicz, and W. Plesiewicz,
Phys. Rev. Lett. \textbf{56}, 508 (1986).

\bibitem{Timm} C. Timm, M. E. Raikh, and F. von Oppen, Phys. Rev. Lett. \textbf{94}, 036602 (2005).

\bibitem{Moca} C. P. Moca, B. L. Sheu, N. Samarth, P. Schiffer, B. Janko, and G. Zarand, arXiv:cond-mat./0705.2016 (2007).

\bibitem {Abrahams} E. Abrahams, P. W. Anderson, D. C. Licciardello, and T. V. Ramakrishnan,
Phys. Rev. Lett. \textbf{42}, 673 (1979).

\bibitem{Zarand} G. Zarand, C. P. Moca, and B. Janko, Phys. Rev. Lett.
\textbf{94}, 247202 (2005).

\bibitem{He} H. T. He, C. L. Yang, W. K. Ge, J. N. Wang, X. Dai, and Y. Q. Wang, Appl. Phys. Lett. \textbf{87}, 162506
(2005).

\bibitem{Matsukura} F. Matsukura, M. Sawicki, T. Dietl, D. Chiba, and H. Ohno, Physica E \textbf{21}, 1032 (2004).

\bibitem{Esch} A. Van Esch, Phys. Rev. B \textbf{56}, 13103 (1997).

\bibitem{Honolka} J. Honolka, S. Masmanidis, H. X. Tang, D. D. Awschalom, and M. L. Roukes,
Phys. Rev. B \textbf{75}, 245310 (2007).

\bibitem{Lee} P. A. Lee and T. V. Ramakrishnan , Rev. Mod. Phys.
\textbf{57}, 287 (1985).

\bibitem{Reinwald} M. Reinwald, U. Wurstbauer, M. D\"{o}ppe, W. Kipferl, K. Wagenhuber,
H.-P. Tranitz, D. Weiss and W. Wegscheider, J. Cryst. Growth
\textbf{278}, 690 (2005).

\bibitem{Nagaev} E. L. Nagaev, Phys. Rev. B \textbf{58}, 816 (1998).

\bibitem{Brands} M. Brands, A. Carl, O. Posth, and G. Dumpich, Phys. Rev. B
\textbf{72}, 085457 (2005).

\bibitem{Brands2} M. Brands, C. Hassel, A. Carl, and G. Dumpich, Phys. Rev. B \textbf{74}, 033406 (2006).

\bibitem{Ono} T. Ono, Y. Ooka, S. Kasai, H. Miyajima, K. Mibu and T. Shinjo, J. Magn. Magn. Mater. \textbf{226}, 1831
(2001).

\bibitem{WAL} D. Neumaier, K. Wagner, S. Gei{\ss}ler, U. Wurstbauer, J. Sadowski, W. Wegscheider, and D. Weiss,
Phys. Rev. Lett. \textbf{99}, 116803 (2007).

\bibitem{Edmonds} K. W. Edmonds, P. Boguslawski, K. Y. Wang, R. P. Campion, S. N. Novikov,
N. R. S. Farley, B. L. Gallagher, C. T. Foxon, M. Sawicki, T. Dietl,
M. Buongiorno Nardelli, and J. Bernholc, Phys. Rev. Lett.
\textbf{92}, 037201 (2004).

\bibitem{Maliepaard} M. C. Maliepaard, M. Pepper, R. Newburry, and G. Hill, Phys. Rev. Lett. \textbf{61}, 369
(1988).


\end{thebibliography}
\end{document}